\documentclass{article}

\usepackage{PRIMEarxiv}

\usepackage[utf8]{inputenc} 
\usepackage[T1]{fontenc}    
\usepackage{hyperref}       
\usepackage{url}            
\usepackage{booktabs}       
\usepackage{amsfonts}       
\usepackage{nicefrac}       
\usepackage{microtype}      
\usepackage{lipsum}
\usepackage{fancyhdr}       
\usepackage{graphicx}       
\graphicspath{{media/}}     

\title{Utilizing Language Models for Tour Itinerary Recommendation}

\author{
  Ngai Lam Ho, Kwan Hui Lim  \\
  Information Systems Technology and Design Pillar \\
  Singapore University of Technology and Design\\
  8 Somapah Road, Singapore 487372 \\
  \texttt{
    ngailam\_ho@mymail.sutd.edu.sg, 
    kwanhui\_lim@sutd.edu.sg} \\
}

\begin{document}
\maketitle

\begin{abstract}
Tour itinerary recommendation involves planning a sequence of relevant Point-of-Interest (POIs), which
combines challenges from the fields of both Operations Research~(OR) and Recommendation Systems
(RS). As an OR problem, there is the need to maximize a certain utility (e.g., popularity of POIs in the
tour) while adhering to some constraints (e.g., maximum time for the tour). As an RS problem, it is heavily
related to problem or filtering or ranking a subset of POIs that are relevant to a user and recommending it
as part of an itinerary. In this paper, we explore the use of language models for the task of tour itinerary
recommendation and planning. This task has the unique requirement of recommending personalized
POIs relevant to users and planning these POIs as an itinerary that satisfies various constraints. We
discuss some approaches in this area, such as using word embedding techniques like Word2Vec and GloVe
for learning POI embeddings and transformer-based techniques like BERT for generating itineraries.
\end{abstract}



\section{Introduction}

The tour itinerary recommendation is a popular and challenging problem, with significant impact for tourism and other domains such as transportation and logistics~\cite{lim2019tour}. The tour itinerary recommendation problem has garnered immense interest in both academia and industry. This problem contains both aspects of an recommendation problem as well as a planning problem. From the recommendation perspective, there are elements of top-K item recommendation and learning to rank, where we aim to recommend a subset of most relevant POIs to a user in the form of an itinerary. From the operation research perspective, it is akin to a constrained optimization problem, where we need to maximize the utility that a user obtains from the planned itinerary while ensuring that the itinerary adheres to certain time and location constraints.

\begin{figure}[t]
    \centering
    \includegraphics[width=\linewidth, trim=0mm 0mm 0mm 0mm, clip]{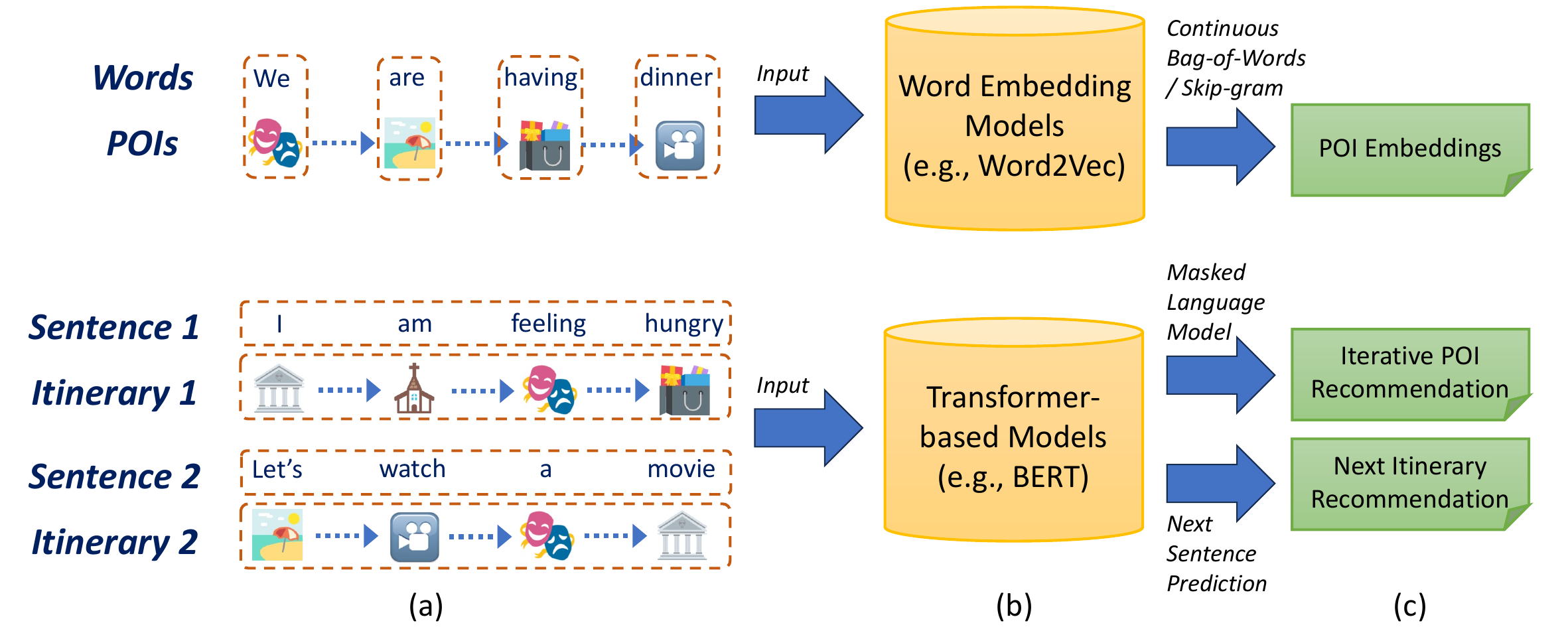}
    \caption{Application of Language Models for Tour Recommendation, including (a) Representation of POIs and Itineraries as Words and Sentences; (b) Feeding of inputs in the form of (a) into various language models; (c) Potential outputs and downstream applications.}
    \label{framework}
\end{figure}

In this paper, we discuss the tour itinerary recommendation problem from both perspectives of a recommendation and operation research problem, particularly on how recent advances on natural language processing have advanced research in this area. In particular, we discuss how language models have been adapted for the task of tour itinerary recommendation such as how word embedding techniques like Word2Vec and GloVe are used for POI representation learning and transformer-based models like BERT are used for next POI and itinerary recommendation. Figure~\ref{framework} provides an overview of potential applications of such a framework.

\section{Tour Itinerary Recommendation Problem}

There are various variants of the tour itinerary recommendation and a popular formulation is based on the Orienteering problem~\cite{golden1987orienteering,gunawan2016orienteering}, which we briefly discuss.

Given a set of POIs $P$, our main objective is to plan for a tour itinerary $T = (p_1,...,p_N)$ that:

\begin{equation}
Max \sum\limits_{p_i}^{P} \sum\limits_{p_j}^P x_{p_i,p_j} S_u(p_i)
\label{eqnMax}
\end{equation} 

{\noindent where $x_{p_i,p_j} = 1$ is the itinerary includes a travelling path from POI $p_i$ to $p_j$, and $x_{p_i,p_j} = 0$ otherwise. $S_u(p_i)$ is the utility score that a user $u$ benefits from based on this tour itinerary $T$.}

Variants of this problem differ in various aspects, mainly in terms of the definition of the utility score used and the types of constraints being implemented. For example, the utility score is typically a global score such as POI popularity for the OR community, while a more personalized interest or relevance score is used for the RS community. Similarly, the two communities might differ in terms of the type and range of constraints from simple ones such as having a fixed number of POIs in a tour itinerary to more diverse constraints such as the need to start/end the tour itinerary at specific locations, complete the tour itinerary within specific time limits or budget constraints, and other unique considerations.

\section{Language Models and Tour Recommendation}

Natural Language Processing (NLP) and particularly large language models have seen rapid progress in recent years with developments ranging from word embedding techniques like Word2Vec~\cite{mikolov2013distributed} and GloVe~\cite{pennington2014glove} to the more recent foundation models like BERT~\cite{devlin2019bert}, GPT~\cite{brown2020language} and their variants. 

Word embedding techniques aim to find appropriate dense vector representation at the word level, by using techniques such as the Continuous Bag of Words and Skip-gram Models that are used in Word2Vec. Both model uses a simple neural network with a single hidden layer and differ in terms of the task. Continuous Bag of Words is used to predict a single target word given the context of its neighbouring words, where Skip-gram does the opposite and tries to predict the neighbouring words that surround a given input word. Thereafter, the weights learned by the hidden layer are used as the word embedding for the specific word, where the number of neurons/nodes in the hidden layer corresponds to the dimension size of the word embedding.

Foundation models like BERT and GPT utilizes the transformer architecture with self-attention and are now commonly used for many downstream NLP tasks from standard text classification to text generation. Models like BERT and its variants uses a self-supervised approach to training and typically use the Mask Language Modelling and Next Sentence Prediction tasks on top of the Transformer architecture. Given a sentence, the Mask Language Modelling task involves hiding a subset of words in the sentence and training the model to predict those words. For Next Sentence Prediction, it is to predict the following sentence given the earlier sentence.

{\bf POI Embedding Models.} Word embedding models have increasingly been used for generating POI embedding. To adapt language models for POI representation learning, the set of POIs in a city can be treated as the vocabulary of words where each POI is akin to an individual word or token. Similar to sentences in NLP, past itineraries or sequences of POI visits are used as a proxy of sentences made up of a series of words. Thereafter, models such as Continuous Bag of Words and Skip-gram are used to learn the vector representation of POIs. Similar models and variants that are adapted for learning latent attributes relating to geographic and temporal factors have been proposed in recent years~\cite{feng2017poi2vec,chang2018content,ho2021user}.

{\bf Transformer-based Models.} More recently, Transformer-based models have gained popularity for not just NLP tasks but increasingly for various types of next POI prediction and tour itinerary recommendation tasks. Similarly, past itineraries comprising sequences of POI visits can be modelled in the same context as sentences that are made up of a series of words. For example, \cite{ho2022poibert} used the BERT model trained on past POI visit sequences, coupled with an iterative process for generating intermediate POIs for recommending tour itineraries. Others like \cite{halder2022poi} have used Transformers for the POI recommendation problem with joint training on the task of next POI prediction and queuing time prediction. More generally, Transformer-based architectures have seen numerous applications for various sequence-related recommendation tasks~\cite{sun2019bert4rec,wu2020sse,chen2019behavior}.

\section{Conclusion}
In this paper, we discussed the problem of tour itinerary recommendation and highlighted its relation to both fields of OR and RS in its consideration of trip constraints and user-relevant recommendation. Following which, we provided an overview of popular techniques used in a variety of NLP tasks and discussed how these NLP techniques, such as word embedding and Transformers, have been adapted to the tour itinerary recommendation task.

\vspace{5mm}
{\noindent {\bf Acknowledgments}. This research is funded in part by the Singapore University of Technology and Design under grant RS-MEFAI-00005-R0201.}


\bibliographystyle{unsrt}  
\bibliography{mybib}

\end{document}